# Chemically Stable Group IV−V Transition Metal Carbide Thin Films in Hydrogen Radical Environments

Abdul Rehman,* Robbert W.E. van de Kruijs, Wesley T.E. van den Beld, Jacobus M. Sturm, and Marcelo Ackermann

Read Online

ACCESS | Metrics & More | Article Recommendations | Supporting Information

**ABSTRACT:** Hydrogen is a crucial element in the green energy transition. However, its tendency to react with and diffuse into surrounding materials poses a significant challenge. Therefore, developing coatings to protect system components in hydrogen environments (molecular, radicals (H*), and plasma) is essential. In this work, we report group IV−V transition metal carbide (TMC) thin films as potential candidates for protective coatings in H* environments at elevated temperatures. We expose TiC, ZrC, HfC, VC, NbC, TaC, and $Co_2C$ thin films, with native surface oxycarbides/oxides ($TMO_xC_y$/$TMO_x$), to H* at elevated temperatures. Based on X-ray photoelectron spectroscopy performed on the samples before and after H*-exposure, we identify three classes of TMCs. HfC, ZrC, TiC, TaC, NbC, and VC (class A) are found to have a stable carbidic-C (TM-C) content, with a further subdivision into partial (class A1: HfC, ZrC, and TiC) and strong (class A2: TaC, NbC, and VC) surface deoxidation. In contrast to class A, a strong carbide reduction is observed in $Co_2C$ (class B), along with a strong surface deoxidation. The H* interaction with TMC/$TMO_xC_y$/$TMO_x$ is hypothesized to entail three processes: (i) hydrogenation of surface C/O atoms, (ii) formation of $CH_x$/$OH_x$ species, and (iii) subsurface C/O atom diffusion to the surface vacancies. The number of adsorbed H atoms required to form $CH_x$/$OH_x$ species (i) and the corresponding thermodynamic energy barriers (ii) are estimated based on the change in the Gibbs free energy ($\Delta G$) for the reduction reactions of TMCs and $TMO_x$. Hydrogenation of surface carbidic-C atoms is proposed to limit the reduction of TMCs, whereas the deoxidation of TMC surfaces is governed by the thermodynamic energy barrier for forming $H_2O$.

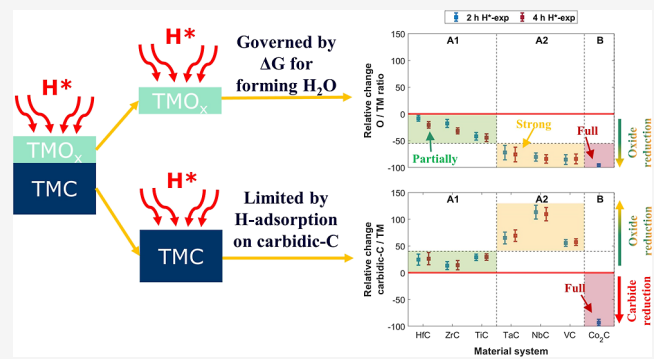

## INTRODUCTION

Hydrogen plays an essential role in a multitude of applications and processes, varying from green energy production to semiconductor manufacturing. For instance, molecular hydrogen is used as fuel in fusion reactors (which is further converted into plasma),[1] hydrogen fuel cells,[2] and aerospace propulsion systems.[3] In the semiconductor industry, hydrogen radicals (H*) are utilized as an etchant and reducing agent, for instance, for removing excess Si for fabricating nanostructures,[4] etching Sn contamination, reducing $RuO_x$ in extreme ultraviolet (EUV) scanners,[5,6] and reducing the surface oxides of group III−V semiconductors.[7] While the high reactivity and small molecular weight of hydrogen make it a promising candidate for many applications, the inclination of hydrogen to react with and diffuse into the walls/components of the systems makes it difficult to work with.[8−11] Hence, to fully exploit the potential of hydrogen, it is necessary to study novel coatings that can endure hydrogen environments and thereby protect hydrogen-sensitive layers.

Group IV−VI transition metal carbides (TMCs) typically exhibit high thermal stability and good molecular hydrogen permeation barrier performance.[12−15] Furthermore, literature indicates that group IV−V TMCs are stable in molecular hydrogen ($H_2$) environments.[16,17] As such, TMCs may be interesting candidates for protective coatings in harsher environments containing chemically active H* and H-ions at increased temperatures, for instance, in fusion reactors and EUV scanners.[11,18] To evaluate the feasibility of TMCs as protective coatings, it is necessary to first assess their chemical stability (reducibility), i.e., the removal of C atoms bonded with transition metal (TM) atoms (carbidic-C), in low-energy H* environments. Such conditions are most relevant to semiconductor operations.[4,7] Understanding the H*-TMC interaction can then help in comprehending TMC reducibility

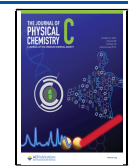

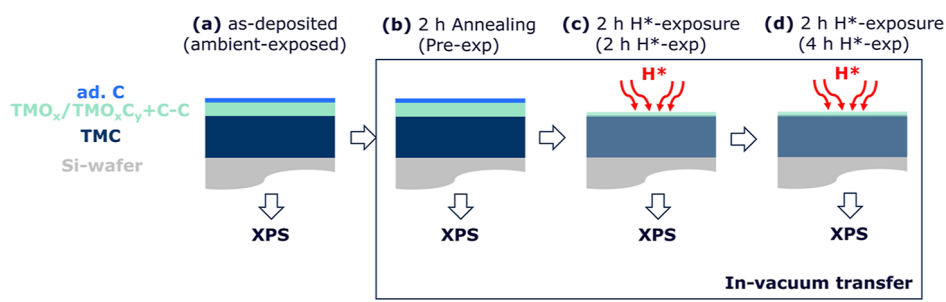

Figure 1. Schematic of methodology. (a) TMC samples were deposited via cosputtering of TM and C targets. Thin $TMO_x/TMO_xC_y$ + C−C and ad. C layers were formed on the samples' surfaces during ambient storage. (b) TMC samples were annealed at elevated temperatures. (c,d) After that, samples were exposed to H* twice. Samples were transferred between the annealing/H*-exposure and AR-XPS chambers under a vacuum pressure of low $10^{-9}$ mbar.

in more energetic and complex hydrogen environments, such as hydrogen plasma.

In recent publications,[19,20] we modeled the reduction (denitridation) of transition metal nitrides (TMNs) in H* in three steps: (i) hydrogenation, (ii) formation of volatile species, and (iii) diffusion of subsurface atoms to the reduced surface. We hypothesize that the reduction (removal of carbidic-C) of TMCs in H* entails similar steps.

(i) Hydrogenation on TMCs predominantly occurs on surface carbidic-C atoms.[12,21] However, the H-adsorption energy on surface carbidic-C atoms decreases as H-loading increases. This limits the maximum H-coverage (H:C) of surface carbidic-C atoms. When each surface carbidic-C atom is occupied by 1 H atom (H:C ratio = 1), further H-adsorption on the surface carbidic-C atom becomes energetically unfavorable.[21]

(ii) Adsorption of sufficient H atoms on a surface carbidic-C atom leads to the formation of volatile $CH_x$ species. The potential formation of volatile $CH_x$ species is related to the change in the Gibbs free energy ($\Delta G$) for the reduction reaction of TMCs (TMC + $x$H → TM + $CH_x$). For instance, $\Delta G$ for the formation of $CH_3$ and $CH_4$ on VC (a candidate for hydrogen-protective coatings[22]) is negative in H* at 1000 K and 0.02 mbar[23,24] (exposure conditions are relevant to EUV scanners and fusion reactors[10,11,25]). Additionally, $\Delta G$ for forming $CH_2$ on $Co_2C$ in H* at 500 K (lower temperature chosen since $Co_2C$ thermally decomposes at 573−623 K[26]) and 0.02 mbar is close to the thermodynamic equilibrium point (Gibbs free energy of $Co_2C$ at 500 K is approximated from its Gibbs formation energy[26]). This suggests that VC is likely to undergo reduction if a surface carbidic-C atom adsorbs 3−4 H atoms, while for reducing $Co_2C$, only 2 H atoms per surface carbidic-C atom are required.

(iii) The formation of volatile $CH_x$ species leads to the formation of C-vacancies on the surface. In order for the reduction reaction to proceed, subsurface C atoms have to diffuse to these surface vacancies. Since the system of TMC/reduced TMC is likely similar to TMC/TM systems, where diffusion of C atoms from the TMC layer to the adjacent TM layer is reported at elevated temperatures,[27−29] we expect that subsurface C atoms will diffuse to the reduced TMC surface at elevated temperatures.

From calculations and the aforementioned literature, (ii) the formation of $CH_x$ species and (iii) the diffusion of subsurface carbidic-C atoms to the reduced TMC surface are shown to be energetically feasible processes at elevated temperatures. However, (i) the number of H atoms that can adsorb on a surface carbidic-C atom (hydrogenation) is expected to be the only factor that could limit the reduction of TMCs. Therefore, we hypothesize that the reducibility of TMCs in H* environments at elevated temperatures is likely governed by their hydrogenation, similar to what was observed during the reduction of TMNs.[19,20] In the following sections, we therefore discuss the thermodynamic feasibility of forming $CH_x$ ($x$ = 1−4) on the studied TMCs, which reflects the number of H atoms that must adsorb per surface carbidic-C atom to form volatile species.

It should be noted that TMCs are metastable in ambient conditions and typically form an approximately 2−5 nm-thick layer of transition metal oxycarbides/oxides ($TMO_xC_y$/$TMO_x$), along with noncarbidic-C (C−C) on the surface.[30] Noncarbidic-C may also form in TMC thin films during the deposition process depending on the deposition conditions.[31−35] Therefore, to comprehend the reducibility of TMCs, the interaction of H* with the surface $TMO_xC_y$/$TMO_x$ and noncarbidic-C must also be taken into account.

The reduction of $TMO_xC_y$/$TMO_x$ is hypothesized to follow the same steps as in the reduction of TMCs and TMNs,[19,20] with the difference that volatile $H_2O$ is formed during the reducing $TMO_xC_y$/$TMO_x$. The interaction of H* with noncarbidic-C mainly depends on its hybridization state ($sp^3$-C and $sp^2$-C).[36,37] We expect that the rate of chemical erosion of noncarbidic-C in TMCs will depend on the type of C and defect sites.[38−40]

To understand the interaction of H* with ambient-exposed TMCs, we expose 5 ± 0.5 nm TiC, ZrC, HfC, VC, NbC, TaC, and $Co_2C$ thin films, with ≈1−2 nm-thick surface $TMO_xC_y$/$TMO_x$ layers, to H*. We perform X-ray photoelectron spectroscopy (XPS) on the samples before and after H*-exposure to evaluate the H*-induced changes in their chemical composition. The behavior of TMCs upon H*-exposure is divided into two main classes based on the observed variation in the carbidic-C fraction. A further subdivision is recognized based on the deoxidation of surface $TMO_xC_y$/$TMO_x$ during H*-exposure. We explain the observed interaction based on the $\Delta G$ calculations for the reduction reactions involved.

## ■ METHODOLOGY

TMC thin films were deposited onto diced Si(100) wafers via DC cosputtering using TM and C targets. The base pressure of the deposition chamber before each deposition is in the lower range of $10^{-8}$ mbar. Ar (99.999%) with a flow rate of 25 sccm was used as the sputtering gas. Working pressure during the deposition was measured to be approximately $9 \times 10^{-4}$ mbar. The deposition rates of TM and C were first calibrated as a function of magnetron currents. For calibration, approximately 20 nm-thick films were deposited. The thickness of the deposited layers was measured via X-ray reflectometry (XRR). The XRR measurements are performed via a Malvern Panalytical Empyrean laboratory diffractometer, which uses a monochromatic Cu-K$\alpha$1 radiation source. Based on the calibration, the magnetron currents were adjusted to achieve a ratio of 1:1 (TM:C) for group IV–V TMCs and 2:1 (TM:C) for Co$_2$C. 5 ± 0.5 nm-thick TMC films were deposited at a deposition rate of approximately 0.07 ± 0.01 nm/s for the study. The thickness is chosen such that the entire depth of the TMC layer is probed using angle-resolved XPS (AR-XPS). The AR-XPS measurements are performed via a Thermo Fisher Theta Probe angle-resolved X-ray photoelectron spectrometer, which uses a monochromatic Al-K$\alpha$ radiation source. During the measurements, spectra were collected at takeoff angles ($\theta$) ranging from 26.75 to 71.75°, with respect to the surface normal, providing a probing depth ranging from approximately 5–1.5 nm, respectively, with a spot size of ≈400 × 400 $\mu$m.

The as-deposited samples were stored in the ambient for approximately a week before the AR-XPS measurements were performed. The AR-XPS measurements revealed the formation of TMO$_x$C$_y$/TMO$_x$ along with noncarbidic-C and adventitious carbon (ad. C) on the samples' surfaces (Figure 1a). In order to saturate all the thermally induced processes before high-temperature H*-exposures, the samples were annealed in a vacuum chamber for 2 h (Figure 1b) at the same temperature that was subsequently used for H*-exposures (Figure 1c). Note that the same chamber/setup is used for H*-exposures, ensuring consistency of temperature settings. The base pressure of the vacuum chamber is in the lower range of $10^{-8}$ mbar, while the maximum pressure of the chamber during annealing is measured to be in the lower range of $10^{-7}$ mbar. The temperature of the sample is measured via an N-type thermocouple which is clamped on the surface of the sample. After annealing, the samples were cooled down to approximately 100 °C in a vacuum, before they were transferred in vacuo (low $10^{-8}$ mbar) to the XPS chamber. AR-XPS measurements were performed on the annealed samples. The corresponding measurements are referred to as "pre-exposed" (pre-exp) in the text and figures and are used as the reference. These XPS measurements mainly reveal the desorption of surface hydrocarbons that the samples accumulated during ambient storage.

The pre-exposed samples were then transferred in vacuo back to the chamber where they were exposed to H* at elevated temperature (Figure 1c). H* are generated by thermally cracking H$_2$ via a W filament, which is heated to approximately 2000 °C. The samples are positioned approximately 0.05 m from the W filament. The working pressure during the H*-exposure is 0.02 mbar, and the corresponding H* flux on the sample surface is $10^{21\pm1}$ m$^{-2}$s$^{-1}$.[41,42] The samples were exposed to H* for 2 h, providing a H* fluence of $7 \times 10^{24\pm1}$ m$^{-2}$, which is an industry-relevant total exposure flux.[10,18] After H*-exposure, the samples were cooled down to approximately 100 °C and then transferred to the XPS chamber. The corresponding XPS measurements are referred to as "2 h H*-exposed" (2 h H*-exp).

Group IV–V TMCs underwent surface cleaning (i.e., deoxidation of surface TMO$_x$C$_y$/TMO$_x$ and chemical erosion of noncarbidic-C) upon 2 h H*-exposure. This resulted in a lesser attenuation of photoelectrons at the surface level, while subsurface C atoms may have also diffused to the surface O-vacancies. Together, these phenomena caused an apparent increase in TMC fraction in the XPS probing depth. In contrast, for the Co$_2$C sample, the intensity of C 1s spectra dropped to almost the noise range, indicating strong carbide reduction in the sample.

In order to avoid ambiguities regarding the stability of carbidic-C in group IV–V TMC samples, the samples were exposed to H* for an additional 2 h under the same conditions (Figure 1d). The XPS measurements on the samples after 4 h H*-exposure are referred to as "4 h H*-exposed" (4 h H*-exp).

Changes in the stoichiometry of the samples due to H*-exposure(s) were evaluated based on the core-level TM, C 1s, and O 1s XPS spectra. The core-level TM and C 1s spectra of pre- and post-H*-exposed samples taken at $\theta$ = 34.25° are discussed in the main text, while the corresponding fitted peak positions are provided in the Supporting Information (Tables S1−S7). Furthermore, a comparison of spectra (TM, C 1s, O 1s, and Si 2p) taken over the range of $\theta$, before and after H*-exposure(s) for each sample, is also provided in the Supporting Information (Figures S7−S14). To quantify carbidic-C and O loss upon H*-exposure(s), the ratios between at. % of carbidic-C and at. % of TM (carbidic-C/TM) and at. % of O and at. % TM (O/TM) in the samples were calculated over the range of $\theta$. The at. % of carbidic-C in the samples was calculated by considering the cumulative area under the carbidic-C peaks (TMC and TMO$_x$C$_y$) in the C 1s spectra and the corresponding sensitivity factor. The whole area under the O 1s spectra was considered to calculate the O-fraction in the samples. Except for the TaC sample, the at. % of the TM in the samples was also calculated by taking the whole area under the core-level XPS spectra of the TM. Due to overlapping binding energies for the Ta 4f and O 2s spectra, the cumulative area under the main peaks of TaC, TaO$_x$C$_y$/TaO$_x$, and Ta$_2$O$_5$ doublets along with the corresponding sensitivity factor was considered for calculating the Ta-fraction in the TaC sample. Changes in the carbidic-C/TM ratios upon H*-exposure(s) are discussed in the main text, while O/TM ratios are provided in the Supporting Information (Figures S1−S3).

We calculate the potential for forming volatile species (CH$_x$ and H$_2$O) in terms of the $\Delta G$ of the relevant reduction reactions (TMC + xH* → TM + CH$_x$ and TMO$_x$ + 2H* → TMO$_{x-1}$ + H$_2$O). The $\Delta G$ is defined by eq 1

$$\Delta G = \sum \nu_P G_P - \sum \nu_R G_R \qquad (1)$$

where $\nu_P$ and $\nu_R$ are the stoichiometric coefficients of the reactants and products, respectively, and $G_P$ and $G_R$ represent the Gibbs free energy ($G$) of the products and reactants, respectively, at the H*-exposure temperature ($T$) and pressure (0.02 mbar).

$G$ ($G_{T,\,0.02\,\text{mbar}}$) at temperature $T$ and pressure 0.02 mbar is given by eq 2

$$G_{T,0.02\text{mbar}} = H_{T,1013.25\text{mbar}} - T(S_{T,1013.25\text{mbar}} + \Delta S) \qquad (2)$$

where $H_{T,1013.25\,\text{mbar}}$ and $S_{T,1013.25\,\text{mbar}}$ are the enthalpy ($H$) and entropy ($S$) at temperature $T$ and standard pressure (1013.25 mbar) in kJ/mol, respectively. For all the studied TMC systems, except $Co_2C$, $H_{T, 1013.25\ mbar}$ and $S_{T, 1013.25\ mbar}$ are obtained from thermodynamic databases.[23,24] For $Co_2C$, G is approximated from the literature.[26] The influence of pressure on $H$ is negligible, especially at low pressures; thus, $H_{T, 0.02\ mbar} \approx H_{T, 1013.25\ mbar}$. Furthermore, the $S$ of solids is also largely unaffected by the pressure changes, meaning $S_{T, 0.02\ mbar} \approx S_{T, 1013.25\ mbar}$ ($\Delta S \approx 0$). However, the $S$ of gases increases as the pressure decreases. The increase in $S$ for ideal gases (where we approximate H*, $CH_x$, $H_2O$, and $H_2$ as ideal gases) due to pressure drop from 1013.25 mbar (standard) to 0.02 mbar (experimental pressure) is given by $R \ln\left(\frac{0.02}{1013.25}\right)$ (where $R$ = ideal gas constant in kJ/mol.K). Hence, in eq 2, $\Delta S = 0.09$ kJ/mol.K for gases (H*, $CH_x$, $H_2O$, and $H_2$) and 0 for solids (TMC, $TMO_x$, and TM).

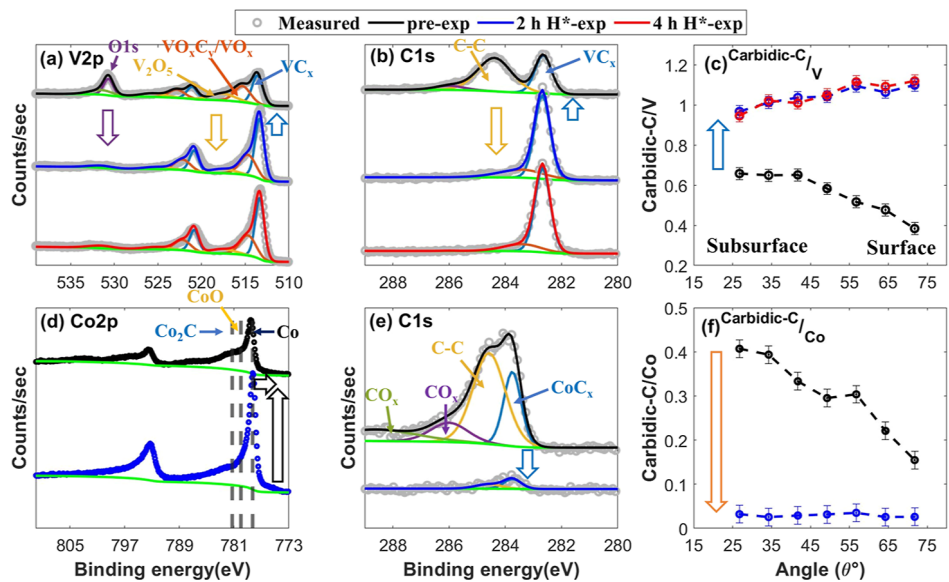

**Figure 2.** XPS core-level TM and C 1s spectra taken at $\theta = 34.25°$, along with carbidic-C/TM ratio as a function of $\theta$ in the pre-exposed (in black), 2 h H*-exposed (in blue), and 4 h H*-exposed (in red) VC (a–c) and $Co_2C$ (d–f) samples. (a) V 2p spectra, (b) C 1s spectra of the VC sample, (c) carbidic-C/V ratio, (d) Co 2p spectra, (e) C 1s spectra of the $Co_2C$ sample, and (f) carbidic-C/Co ratio. Upon 2 h H*-exposure, the VC sample undergoes strong surface deoxidation, resulting in an increase in the VC fraction in the XPS probing depth. The stoichiometry of the sample remains unchanged upon an additional 2 h H*-exposure (4 h H*-exp), indicating that the carbidic fraction in the VC sample is stable in H*. In contrast, the $Co_2C$ sample undergoes significant carbide and oxide reduction upon 2 h H*-exposure.

It is important to note that the $H$ and $S$ values used to calculate the $\Delta G$ for the reduction reactions are based on bulk materials[23,24,26] and may differ for the deposited thin films. Nevertheless, despite potential discrepancies, using bulk reference structures ensures consistency in the $\Delta G$ calculations, facilitating a reliable comparison across the studied TMCs.

## ■ RESULTS AND DISCUSSION

In this section, we present a detailed analysis of our observations in relation to the proposed thermodynamic model. We begin by discussing the interaction of H* with the VC sample at 1000 K and the $Co_2C$ sample at 500 K. Next, we introduce a predictive model for the stability of carbidic-C in HfC, ZrC, TaC, and NbC thin films under H*-exposure at 1000 K. This model also addresses the deoxidation of surface $TMO_xC_y/TMO_x$. We then evaluate the observed interaction of H* with HfC, ZrC, TaC, and NbC thin films, relating it to the model. Finally, we conclude the section with a holistic analysis of the observations, the model, and its inherent limitations.

**H* Interaction with VC and $Co_2C$.** The VC sample predominantly undergoes surface deoxidation upon 2 h H*-exposure. The intensity of the $V_2O_5$ doublet in the V 2p XPS spectra decreased to the noise level (Figure 2a). Analogously, a significant drop in the O/V ratio is also noted (Figure S2c). This results in an increase in VC fraction in the XPS probing depth. This is reflected by an increase in the intensity of $VC_x$ doublet in the V 2p, the $VC_x$ peak in the C 1s, and the carbidic-C/V ratio after 2 h H*-exposure (Figure 2a–c). Further exposure to H* for an additional 2 h (4 h H*-exp) does not change the sample's stoichiometry significantly (Figure 2a–c). This suggests that carbidic-C in the VC sample is stable under the performed experimental conditions.

The $Co_2C$ sample undergoes strong carbide/oxide reduction, in contrast to the stable carbidic-C in VC. The Co 2p spectra are not deconvoluted due to the lack of references on Co° asymmetric characteristics and satellite features of distinct oxidation states of Co. Nevertheless, a significant increase in the intensity of Co 2p at lower binding energies following 2 h H*-exposure indicates a strong reduction of the sample (Figure 2d). Likewise, a significant drop in the intensity of the $CoC_x$ peak in the C 1s spectra and the carbidic-C/Co ratio suggests a strong carbide reduction (Figure 2e,f). Furthermore, a strong reduction of surface oxides is confirmed by the decrease in the O/Co ratio after 2 h H*-exposure (Figure S3).

**Thermodynamic Model.** Thermodynamically, the formation of $CH_3$ and $CH_4$ is feasible on VC under the performed H*-exposure conditions (see Figure 3 for calculations), indicating that the formation of volatile species on VC is energetically feasible (second step in the hypothesis). Furthermore, sufficient carbidic-C atoms are present at the surface level (Figure 2c). Therefore, it is unlikely that the diffusion of subsurface C atoms to the surface is hindering the reduction of the VC sample (third step in the hypothesis). Consequently, based on our hypothesis, the observed stability of carbidic-C in VC (Figure 2a–c) indicates that the adsorption of 3−4 H atoms per surface carbidic-C atom is unfeasible (first step in the hypothesis). This aligns with the observed reduction of $Co_2C$ (Figure 2d,f), where the formation of $CH_2$ is already thermodynamically feasible, in addition to $CH_3$ and $CH_4$ (Figure 3), necessitating the adsorption of only 2 H atoms per surface carbidic-C atom for the reduction reaction. Therefore, based on the observations (Figure 2) and literature,[21] we propose that under the performed experimental conditions, the hydrogenation of TMCs is limited to fewer than 3 H atoms per surface carbidic-C atom. Therefore, hydrogenation of surface carbidic-C atoms limits the reduction of TMCs in H* environments at elevated temperatures.

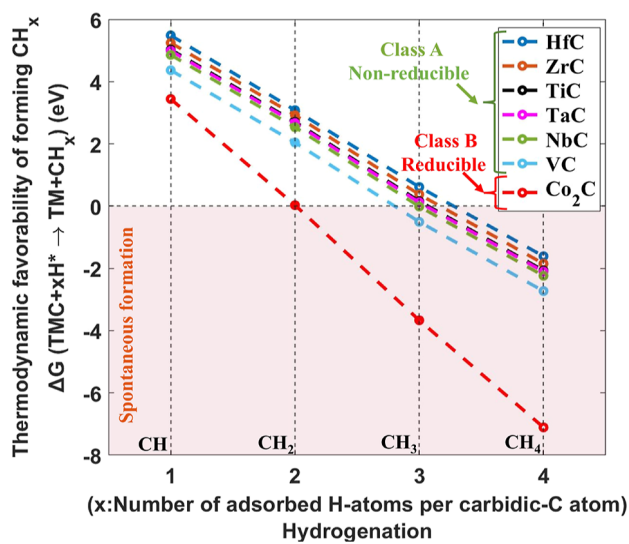

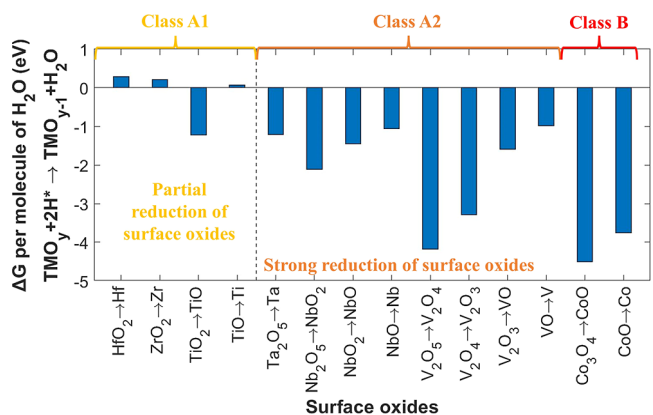

**Figure 3.** $\Delta G$ for the reduction of TMCs per molecule of $CH_x$ ($x = 1$, 2, 3, and 4) calculated at 0.02 mbar and 1000 K (for HfC, ZrC, TiC, TaC, NbC, and VC) and 500 K (for $Co_2C$). Similar to VC, HfC, ZrC, TiC, TaC, and NbC are expected to be nonreducible, as adsorption of 3−4 H atoms per surface carbidic-C atom is energetically unfeasible. These TMCs are referred to as class A. Conversely, $Co_2C$ requires only 2 H atoms per surface carbidic-C atom for reduction, making it reducible under the performed H*-exposure conditions (class B).

**Figure 4.** $\Delta G$ for the reduction of $TMO_x$ per molecule of $H_2O$ calculated at 0.02 mbar and 1000 K (for $HfO_x$, $ZrO_x$, $TiO_x$, $TaO_x$, $NbO_x$, and $VO_x$) and 500 K (for $CoO_x$). $\Delta G$ for the complete reduction of oxides that may form on HfC, ZrC, and TiC in ambient or during H*-exposure is positive. Therefore, surfaces of these TMCs are expected to undergo only partial deoxidation in H* (class A1). In contrast, the complete reduction of $TMO_x$ on the surfaces of TaC, NbC, VC, and $Co_2C$ has a negative $\Delta G$. Hence, surfaces of TaC and NbC are expected to undergo strong deoxidation, similar to the observed behavior in the VC and $Co_2C$ samples. TaC, NbC, and VC are referred to as class A2, whereas $Co_2C$ is referred to as class B, as it undergoes carbide reduction along with strong surface deoxidation.

Deoxidation of surface $TMO_xC_y/TMO_x$ is observed in the VC and $Co_2C$ samples upon H*-exposure (Figures 2a,d, S2c, and S3). The reduction of surface $TMO_xC_y/TMO_x$ is expected to follow the same processes as in the reduction of TMC layers below. Since the thermochemical properties of $TMO_xC_y$ are not available in the literature and the surface layer on ambient-exposed TMCs is typically O-rich, with predominating TM-O bonds, we model the surfaces of TMCs as pristine $TMO_x$. Hydrogenation of surface O atoms and diffusion of subsurface O atoms toward the surface vacancies are energetically feasible processes, as suggested by the observed deoxidation of the samples, also consistent with the literature.[43,44] Nevertheless, the thermodynamic energy barrier for forming volatile species ($H_2O$: second step in the hypothesis) is expected to govern the deoxidation of TMC surfaces. Thus, we propose that the deoxidation of the surfaces of ambient-exposed TMCs in H* is primarily governed by the thermodynamic barrier for forming $H_2O$ on $TMO_x$ (Figure 4).

**Application of the Model on Group IV−V TMCs.** The hydrogenation of surface carbidic-C atoms is proposed to be limited to fewer than 3 H atoms per carbidic-C atom under the performed H*-exposure conditions. $\Delta G$ for the reduction reaction of TMCs indicates that only the formation of $CH_4$ is spontaneous on HfC, ZrC, TiC, and TaC (Figure 3). Thus, 4 H atoms are required to adsorb on a surface carbidic-C atom for the reduction reaction to occur on these TMCs. Similar to VC, $CH_3$ formation is spontaneous on NbC (Figure 3), necessitating 3 H atoms per surface carbidic-C atom for a thermodynamically feasible reduction. Therefore, based on the model, we expect that carbidic-C in HfC, ZrC, TiC, TaC, and NbC is stable, similar to VC. These TMCs are referred to as class A in the text and figures, while $Co_2C$ reduces under the performed experimental conditions due to the thermodynamically feasible formation of $CH_2$, hence it is referred to as class B.

The thermodynamic energy barrier for forming $H_2O$ on TMCs (i.e., the $TMO_xC_y/TMO_x$ top layer) is proposed to govern the deoxidation of TMC surfaces during H*-exposure. For $HfO_2$ and $ZrO_2$, $\Delta G$ for the complete reduction of oxides ($TMO_x \rightarrow TM$) is positive, Furthermore, for $TiO_2$, while $\Delta G$ for the partial reduction to TiO is negative, it remains positive for the full reduction to Ti. Therefore, we expect that the surfaces of HfC, ZrC, and TiC will only undergo partial deoxidation. These TMCs are referred to as class A1. Conversely, $\Delta G$ for both partial and full reduction of oxides on TaC, NbC, VC, and $Co_2C$ is negative, indicating that surface oxides on TaC and NbC are likely to undergo strong deoxidation, similar to the observed strong deoxidation of the VC and $Co_2C$ samples. Consequently, TaC, NbC, and VC samples are classified as class A2. However, $Co_2C$ is referred to as class B due to its unstable carbidic content.

**(Class A1) No Effective Carbide Reduction along with a Partial Surface Deoxidation.** Similar to the VC sample, the HfC, ZrC, and TiC samples also undergo surface deoxidation upon 2 h H*-exposure. However, the extent of deoxidation is less pronounced in these samples compared to what is observed in the VC samples. Notably, the decrease in the intensity of the oxide doublets in the Hf 4f, Zr 3d, and Ti 2p XPS spectra is less significant when compared to the oxide doublet in V 2p (Figures 5a,d,g, and 2a). Furthermore, the TiC sample undergoes stronger surface deoxidation than the HfC and ZrC samples (Figure 5a,d,g). This trend is also evident in the drop of the O/TM ratios upon H*-exposure (Figure S1). The stronger deoxidation of the TiC sample compared to the HfC and ZrC sample is attributed to the thermodynamically feasible partial reduction of $TiO_2$ to TiO, unlike $HfO_2$ and $ZrO_2$, aligning with our proposed model (Figure 4).

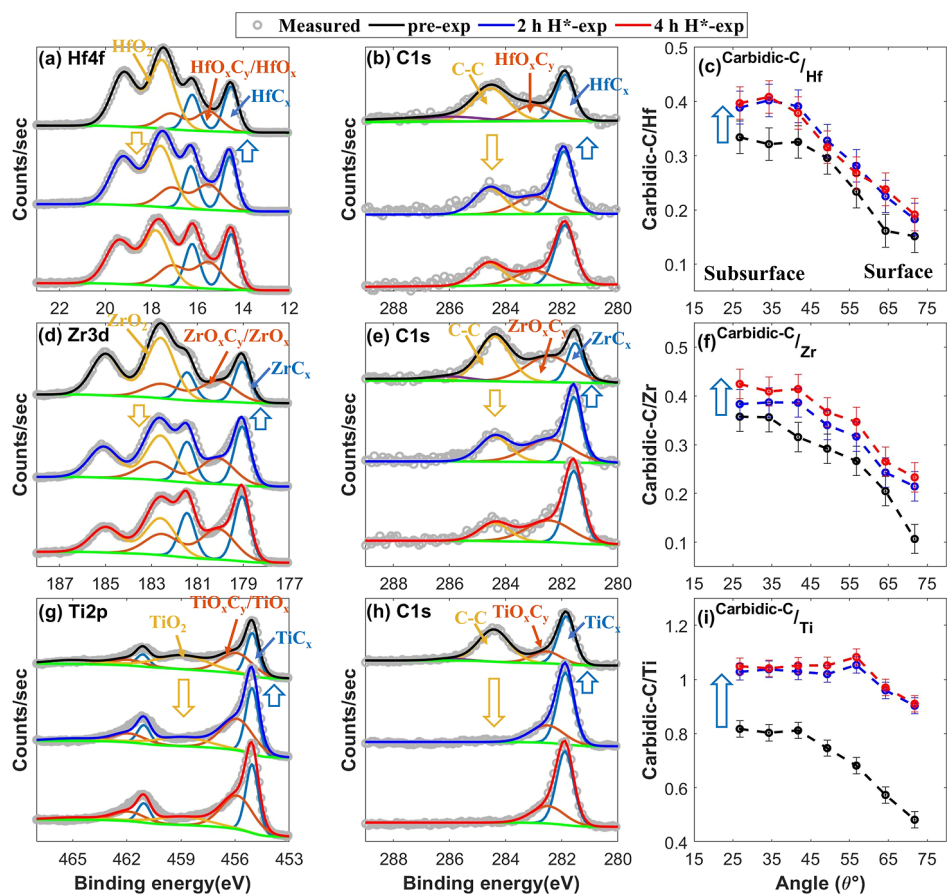

**Figure 5.** XPS core-level TM and C 1s spectra taken at $\theta = 34.25°$ along with carbidic-C/TM ratio as a function of $\theta$ in the pre-exposed (in black), 2 h H*-exposed (in blue), and 4 h H*-exposed (in red) HfC (a–c), ZrC (d–f), and TiC (g–i) samples. (a) Hf 4f spectra, (b) C 1s spectra of the HfC sample, (c) carbidic-C/Hf ratio, (d) Zr 3d spectra, (e) C 1s spectra of the ZrC sample, (f) carbidic-C/Zr ratio, (g) Ti 2p spectra, (h) C 1s spectra of the TiC sample, and (i) carbidic-C/Ti ratio. An increase in the carbidic-C/TM ratio following 2 h H*-exposure is due to the removal of O atoms from the surface. No significant change in the C 1s spectra and carbidic-C/TM ratio after 2 h H*-exposure suggests that the carbidic fraction in the HfC, ZrC, and TiC samples is stable in H*. Moreover, the stronger increase in the TMC fraction in the TiC samples is due to more pronounced surface deoxidation of the TiC sample compared to that in the HfC and ZrC samples.

The removal of O atoms from the surface level results in an increase in the intensity of the TMC doublets in the core-level TM and carbidic-C peaks in the C 1s XPS spectra of the HfC, ZrC, and TiC samples upon 2 h H*-exposure (Figure 5a,b,d,e,g,h). Correspondingly, the carbidic-C/TM ratios also increased (Figure 5c,f,i). Notably, the increase in the carbidic-C/TM ratio is more pronounced in the TiC sample than in the HfC and ZrC samples. This is due to less surface deoxidation in the HfC and ZrC samples than in the TiC sample.

Upon an additional 2 h H*-exposure (4 h H*-exp), no significant change in the chemical composition of the samples is observed (Figure 5). This shows that carbidic-C in the HfC, ZrC, and TiC samples is stable in H*.

Sufficient carbidic-C atoms are present at the surface level after 2 h H*-exposure, and no loss of carbidic-C atoms is observed after 4 h H*-exposure (Figure 5). Furthermore, thermodynamically, $CH_4$ formation is feasible on HfC, ZrC, and TiC (Figure 3). Therefore, consistent with our model, the stability of carbidic-C in HfC, ZrC, and TiC is attributed to the limited hydrogenation of surface carbidic-C atoms, i.e., less than 4 H atoms per carbidic-C atom.

A discrepancy in the chemical erosion of noncarbidic-C (C–C) in the samples upon 2 h H*-exposure is observed (Figure 5b,e,h). This could be due to the difference in the hybridization state ($sp^2$-C and $sp^3$-C) of C–C and the extent of carbide/oxide passivation on C–C.[31,33,35,40] Nevertheless, further investigations are required to understand the H*-induced chemical erosion of the noncarbidic-C phase in TMC systems.

**(Class A2) No Effective Carbide Reduction along with a Strong Surface Deoxidation.** In contrast to the partial surface deoxidation observed on TiC, ZrC, and HfC, surfaces of the TaC and NbC samples undergo strong deoxidation upon 2h H*-exposure, similar to the VC sample (Figures S1 and S2). The intensity of oxide doublets in the TaC and NbC samples dropped to the noise level after 2 h H*-exposure (Figure 6a,d). The strong surface deoxidation of the samples is consistent with our model, where $\Delta G$ values for the complete reduction ($TMO_x \rightarrow TM$) of $TaO_x$ and $NbO_x$ are negative (Figure 4). Accordingly, a stronger surface deoxidation results in a higher increase in the TMC fraction in the XPS probing depth in the TaC and NbC samples than in the HfC, ZrC, and VC samples (Figures 5 and 6).

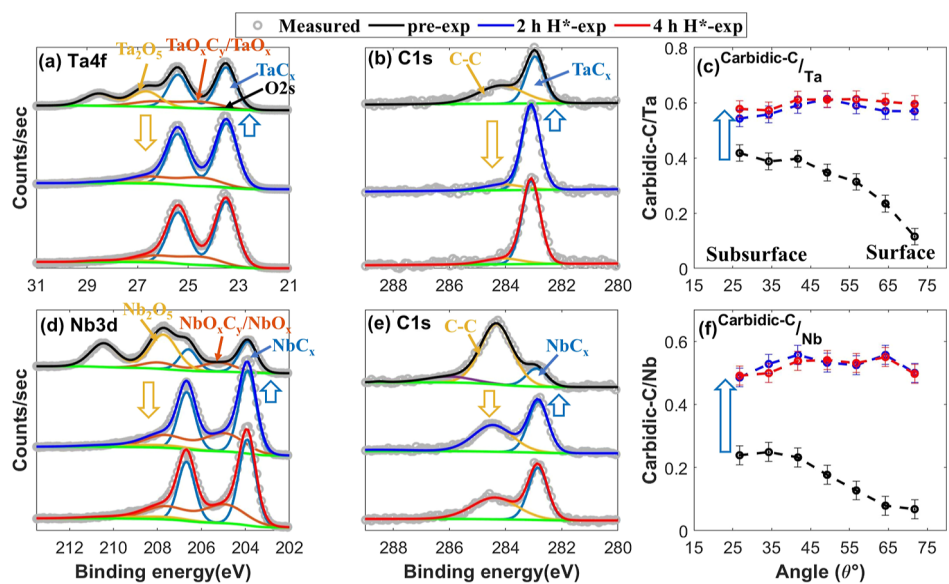

**Figure 6.** XPS core-level TM and C 1s spectra taken at $\theta = 34.25°$, along with carbidic-C/TM ratio as a function of $\theta$ in the pre-exposed (in black), 2 h H*-exposed (in blue), and 4 h H*-exposed (in red) TaC (a–c) and NbC (d–f) samples. (a) Ta 4f spectra, (b) C 1s spectra of the TaC sample, (c) carbidic-C/Ta ratio, (d) Nb 3d, (e) C 1s spectra of the NbC sample, and (f) carbidic-C/Nb ratio. Surfaces of the TaC and NbC samples undergo strong deoxidation upon 2 h H*-exposure, similar to the VC sample, resulting in a strong increase in the TMC fraction in the XPS probing depth. No change in the stoichiometry of the samples is observed after 2 h H*-exposure, indicating that carbidic fraction in the TaC and NbC samples is nonreducible in H*.

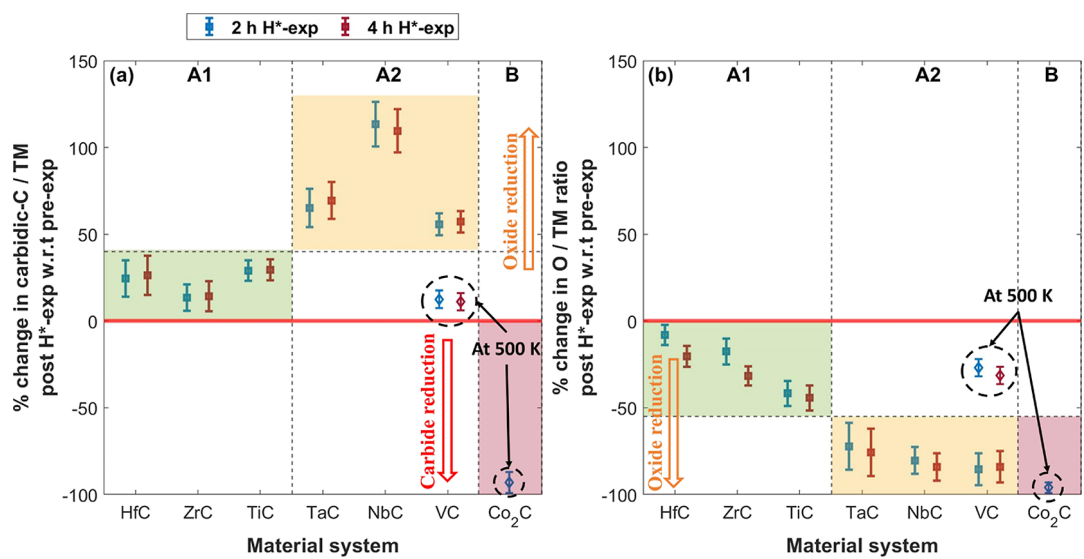

**Figure 7.** Relative change in the post-H*-exposed samples with respect to the pre-exposed samples at $\theta = 34.25°$. (a) Relative change in carbidic-C/TM ratio and (b) relative change in O/TM ratio. TMCs are categorized into three classes. The HfC, ZrC, TiC, TaC, NbC, and VC samples (class A) show no sign of carbide reduction. Surface oxides on the TaC, NbC, and VC samples (class A2) show stronger reduction than on the HfC, ZrC, and TiC samples (class A1) upon 2 h H*-exposure. The drop in carbidic-C/Co and O/Co ratios following 2 h H*-exposure is due to the reduction of carbides and oxides in the $Co_2C$ sample (class B).

No change in the chemical composition of the TaC and NbC samples is noted upon an additional 2 h H*-exposure (4 h H*-exposed) (Figure 6), indicating that the carbidic fraction in the TaC and NbC samples is nonreducible under the performed experimental conditions.

Similar to the HfC, ZrC, TiC, and VC samples, carbidic-C atoms are present at the surface level of the TaC and NbC samples as well (Figure 6c,f). Furthermore, thermodynamically, the formation of volatile species is feasible on TaC and NbC (Figure 3). Therefore, the nonreducibility of TaC and NbC is attributed to the limited hydrogenation of surface carbidic-C atoms. Since the formation of $CH_3$ is feasible on NbC, similar to VC (Figure 3), in line with our model, it suggests that the hydrogenation on these TMCs is limited to less than 3 H atoms per surface carbidic-C atom.

Similar to class A1 TMCs, a discrepancy in the chemical erosion of C−C is noted in the TaC and NbC samples (Figure

6b,e). Therefore, further research is necessary to comprehend the chemical erosion of C−C in TMC systems.

**Validation of the Model.** Based on the relative change in carbidic-C/TM and O/TM ratios in the samples after 2 and 4 h H*-exposure with respect to the pre-exposed samples, we categorize the studied TMCs into three classes (Figure 7).

Class A (HfC, ZrC, TiC, TaC, NbC, and VC) is characterized by an initial increase in carbide (XPS) signals after 2 h H*-exposure due to deoxidation of the surface, followed by no further apparent change upon an additional 2 h H*-exposure (4 h H*-exp) (Figure 7a). This indicates that the carbidic fraction in class A TMCs is stable.

The XPS results indicate that the carbide reduction reaction on class A TMCs is not limited due to the depletion of carbidic-C atoms at the surface (Figures 2c, 5c,f,i, 6c,f). Furthermore, thermodynamically, the formation of $CH_x$ species is feasible for these TMCs (Figure 3). Thus, based on our observations and $\Delta G$ calculations, we propose that the hydrogenation of surface carbidic-C atoms limits the reduction of carbidic content in TMCs, i.e., the hydrogenation is effectively limited to fewer than 3 H atoms per surface carbidic-C atom. This is also consistent with the literature on the hydrogenation of TMCs in $H_2$,[21] where the adsorption of only 1 H atom per surface carbidic-C atom is favorable. Nevertheless, the reported simulations only consider the hydrogenation of model systems in $H_2$ environments. It is plausible that more than 1 H atom could adsorb on a surface carbidic-C atom in a disordered (amorphous or polycrystalline) ambient-exposed TMC in H* environment. This is in line with the observed reduction of $Co_2C$ (class B), where adsorption of at least 2 H atoms per surface carbidic-C atom is required for the reduction to occur.

Furthermore, the thermodynamic calculations performed at a lower temperature of 500 K for $Co_2C$ exhibit a steeper slope (Figure 3). This indicates that the formation of $CH_x$ on TMCs is thermodynamically more favorable at a lower temperature, making the formation of $CH_3$ feasible on all the class A TMCs at 500 K (Figure S4). Since VC is likely the most reducible among class A TMCs, we also expose VC to H* at 500 K. Similar to the results at 1000 K, carbidic-C in the VC sample is stable in H* at 500 K (Figures 7a and S5). However, the surface deoxidation of the sample at 500 K is less pronounced compared to that at 1000 K (Figures 7b, S2c, and S6). This difference could be attributed to limited/constrained diffusion of subsurface C/O atoms to the surface and/or variations in the rate of desorption/readsorption of volatile species. Despite these differences, the stability of the carbidic-C in VC at 500 K, similar to 1000 K, suggests that our proposed model is also valid at lower temperatures; i.e., the hydrogenation of surface carbidic-C atoms is limited to less than 3 H atoms per carbidic-C atom.

Class A TMCs are subdivided into classes A1 (HfC, ZrC, and TiC) and A2 (TaC, NbC, and VC) based on the amount of surface deoxidation during 2 h H*-exposure, which is either small (A1) or large (A2) (Figure 7b). Notably, the HfC and ZrC samples show surface deoxidation despite a positive $\Delta G$ for the reduction reaction of $HfO_2$ and $ZrO_2$. This is attributed to O-loss from surface $TMO_xC_y$ layers, which are not considered due to the lack of thermochemical data. Nevertheless, the overall trend in the observed deoxidation of the samples is consistent with our model, which posits that the surface deoxidation of TMCs is governed by $\Delta G$ for forming $H_2O$ on $TMO_x$ (Figure 4).

Note that the $\Delta G$ calculations presented in this work are based on bulk materials and do not account for potential variations in structural and surface properties specific to our thin-film samples. Nevertheless, these bulk reference calculations provide a robust framework for explaining the observed H*-TMC interactions.

## ■ CONCLUSIONS

Ambient-exposed group IV−V TMC and $Co_2C$ thin-film samples are exposed to H* at elevated temperatures. In the HfC, ZrC, TiC, TaC, NbC, and VC samples, an initial deoxidation of surface oxycarbides/oxides is observed upon H*-exposure. After that, no further change in the chemical composition of the samples is noted. The results show that the carbide fraction in the HfC, ZrC, TiC, TaC, NbC, and VC samples is nonreducible. In contrast, the $Co_2C$ sample showed strong carbide and oxide reduction. Furthermore, a discrepancy in the deoxidation of the TMC surfaces is noted, i.e., the TaC, NbC, VC, and $Co_2C$ samples undergo stronger surface deoxidation than the HfC, ZrC, and TiC samples.

We propose a model wherein the number of H atoms that can adsorb on surface carbidic-C atoms (hydrogenation) limits TMC reduction in H* environments. Based on the $\Delta G$ calculations, HfC, ZrC, TiC, and TaC should undergo reduction by forming $CH_4$, while NbC and VC can also reduce by $CH_3$ formation. Notably, the nonreducible samples' carbidic content indicates that H-adsorption on surface carbidic-C atoms is insufficient to form $CH_4$ or $CH_3$—so fewer than 3 H atoms adsorb per surface carbidic-C atom. Furthermore, the observed reduction of $Co_2C$ aligns with our model, as it can reduce by forming $CH_2$, necessitating only 2 H atoms per surface carbidic-C atom to form volatile species.

In addition to that, we propose that the deoxidation of TMC surfaces is governed by the thermodynamic barrier for forming $H_2O$ on $TMO_x$. Since $\Delta G$ for complete reduction of $TaO_x$, $NbO_x$, $VO_x$, and $CoO_x$ is negative, surfaces of the TaC, NbC, VC, and $Co_2C$ samples undergo strong deoxidation. Whereas thermodynamically complete reduction of $HfO_x$, $ZrO_x$, and $TiO_x$ is unfeasible, hence, surfaces of the HfC, ZrC, and TiC samples show only partial deoxidation.

The results indicate that group IV−V TMCs are chemically stable in H* at elevated temperatures, making them potential candidates for protective coatings in H* environments. Moreover, the study also highlights that surface cleaning (deoxidation of the surface and chemical erosion of non-carbidic-C) of TMCs occurs during H*-exposure. This suggests that H* could potentially be used to enhance or maintain the surface response of group IV−V TMCs, which is relevant for catalytic applications.

## ■ ASSOCIATED CONTENT

### ⓈSupporting Information

The Supporting Information is available free of charge at https://pubs.acs.org/doi/10.1021/acs.jpcc.4c04822.

> O/TM ratios; fitted XPS spectra of the VC sample exposed to H* at 500 K; $\Delta G$ for the formation of $CH_x$ on the studied TMCs calculated at 500 K; comparison between the core-level TM, C 1s, O 1s, and Si 2p XPS spectra of the pre- and post-H*-exposed samples as a function of $\theta$; and fitted XPS peak positions in the core-level TM and C 1s XPS spectra of the pre- and post-H*-exposed sample taken at $\theta = 34.25°$ (PDF)

■ AUTHOR INFORMATION

**Corresponding Author**

Abdul Rehman − *Industrial Focus Group XUV Optics, MESA + Institute for Nanotechnology, University of Twente, Enschede 7522NB, The Netherlands*; orcid.org/0000-0002-8846-9975; Email: a.rehman@utwente.nl

**Authors**

Robbert W.E. van de Kruijs − *Industrial Focus Group XUV Optics, MESA+ Institute for Nanotechnology, University of Twente, Enschede 7522NB, The Netherlands*; orcid.org/0000-0002-4738-0819

Wesley T.E. van den Beld − *Industrial Focus Group XUV Optics, MESA+ Institute for Nanotechnology, University of Twente, Enschede 7522NB, The Netherlands*; orcid.org/0000-0002-5449-3838

Jacobus M. Sturm − *Industrial Focus Group XUV Optics, MESA+ Institute for Nanotechnology, University of Twente, Enschede 7522NB, The Netherlands*; orcid.org/0000-0002-0731-6329

Marcelo Ackermann − *Industrial Focus Group XUV Optics, MESA+ Institute for Nanotechnology, University of Twente, Enschede 7522NB, The Netherlands*

Complete contact information is available at:
https://pubs.acs.org/10.1021/acs.jpcc.4c04822

**Notes**

The authors declare no competing financial interest.

■ ACKNOWLEDGMENTS

This work has been carried out in the frame of the Industrial Partnership Program "X-tools," Project No. 741.018.301, funded by The Netherlands Organization for Scientific Research, ASML, Carl Zeiss SMT, and Malvern Panalytical. We acknowledge the support of the Industrial Focus Group XUV Optics at the MESA + Institute for Nanotechnology at the University of Twente.